\documentclass[submission,copyright,creativecommons]{eptcs}
\providecommand{\event}{GASCom 2024} % Name of the event you are submitting to

\usepackage{iftex}
\usepackage{graphicx} % Required for inserting images
\usepackage{amsmath}
\usepackage{amsfonts}
\usepackage{amsthm}
\usepackage{algorithm}
\usepackage[noend]{algpseudocode}
\usepackage{hyperref}
\usepackage{array}
\usepackage{appendix}
\usepackage{xcolor}
\usepackage{orcidlink}

\newtheorem{lemma}{Lemma} % define a new lemma environment
\newtheorem{theorem}{Theorem}

\ifpdf
  \usepackage{underscore}         % Only needed if you use pdflatex.
  \usepackage[T1]{fontenc}        % Recommended with pdflatex
\else
  \usepackage{breakurl}           % Not needed if you use pdflatex only.
\fi

\title{Optimal Generation of Strictly \\ Increasing Binary Trees and Beyond}
\author{Olivier Bodini\orcidlink{0000-0002-1867-667X}
\institute{LIPN\\ Villetaneuse, France}
\institute{Institut Galilé\\
Université Sorbonne Paris-Nord}
\email{olivier.bodini@lipn.univ-paris13.fr}
\and
Francis Durand\orcidlink{0009-0004-4146-0289}
\institute{LIPN\\ Villetaneuse, France}
\institute{Institut Galilé\\
Université Sorbonne Paris-Nord}
\email{francis.durand@ens-paris-saclay.fr}
\and
Philippe Marchal\orcidlink{0000-0001-8236-5713}
\institute{LAGA\\ Villetaneuse, France}
\institute{Institut Galilé\\
Université Sorbonne Paris-Nord}
\email{marchal@math.univ-paris13.fr}}
\def\titlerunning{Generation of Strictly Increasing Binary Trees}
\def\authorrunning{O. Bodini, F. Durand \& P. Marchal}
\begin{document}
\maketitle

\section{Introduction}

Tree-like data structures are fundamental in computer science, serving as critical tools for modeling phenomena, testing, and creating diverse and representative datasets that enable effective training of machine learning frameworks by exposing models to a wide range of possible input scenarios. Recognizing the necessity for random samplers of these structures, a comprehensive body of work has emerged offering algorithms for generating simple tree families. These algorithms, ranging from general approaches like the Recursive method and Boltzmann sampler \cite{DFLS004} to specific ones like the BBJ algorithm for binary trees \cite{BBJ17}, excel in efficiency and minimal randomness usage. Despite the abundance of techniques for simple trees, the random generation of increasing trees—a vital component in priority queue management—remains underexplored. This gap is attributed to their non-uniform internal structure, challenging the creation of homogeneous algorithms. 
Our paper introduces a groundbreaking algorithm for the optimal generation of strictly increasing binary trees, leveraging a novel approach pioneered by Ph. Marchal \cite{Marchal2012} that ensures both entropy and time efficiency and we prove here random bit complexity. Additionally, we present an enhanced algorithm that adopts an innovative approximation schema for the recursive method. This method is tailored for all weighted unary-binary increasing trees, guaranteeing minimal randomness consumption. This dual approach not only advances the field of random increasing tree generation but also sets a new standard for algorithmic efficiency and randomness optimization.
The next section is devoted to presenting the first algorithm and evaluating its cost. Section 3 addresses the second algorithm, which represents an improvement of the recursive method through the use of a Monte Carlo process. The final section demonstrates how the second method can be adapted to operate within a more general framework.

\section{An Ad Hoc Approach for Sampling Strictly Increasing Trees}

Before introducing the first algorithm, let us make a first easy but important observation. Strictly increasing binary trees are intimately related to permutations (this has been known for a long time). Given a sequence of distinct real numbers in the interval [0,1], we can derive a permutation based on the relative order of the sequence, and thus an increasing tree in the following way. Starting with the position of the smallest number in the sequence, we place a 1 at this position in the permutation (and at the same time, we create a root for the tree), and continue this process until all the numbers have been placed (in the permutation and in the binary tree). If the sequence alternates between increasing and decreasing values, the resulting permutation will also alternate and the tree will be a strictly increasing binary tree. Furthermore, suppose the sequence is chosen randomly and uniformly. In that case, the permutation reflects a uniform distribution of all possible permutations and the binary tree constructed is uniform among binary trees of size $n$. This leads us to look for an algorithm that generates a random, uniform, and alternating sequence, which in turn allows us to construct a uniform alternating permutation and hence a uniform strictly increasing binary tree.\\

\begin{algorithm}[h!]
\caption{Generate an Alternating Permutation of Size $n$}
\begin{algorithmic}[1]
\Require $n > 1$
\State $D, Y \gets \text{arrays of size } n$
\State $p \gets \lfloor\log_2(n)\rfloor+1$
\While{precision on $Y$ is not sufficient}
    \State $p \gets p+1 $
    \While{$Y$ is empty}
        \State $r \gets U_0 \text{with } p \text{ digits.}$
        \State $D[0] \gets 1 - \sin^2\left(\frac{\pi}{2} r\right)$
        \For{$i \gets 0$ \textbf{to} $n - 2$}
            \State $r \gets U_{i+1} \text{with } p \text{ digits.}$
            \State $D[i + 1] = r^2(1 - D[i])$
        \EndFor
        \State $\alpha \gets \sqrt{\frac{1 - D[n - 1]}{1 - D[0]}}$
        \State $threshold \gets \frac{1}{\alpha + \alpha^{-1}}$
        \State $proba \gets \text{rand()}$
        \If{$proba \leq threshold$}
            \State $Y \gets (1 - D[0], D[1], 1 - D[2], \ldots)$
        \ElsIf{$threshold < proba \leq 2 \cdot threshold$}
            \State $Y \gets (1 - D[n - 1], D[n - 2], 1 - D[n - 3], \ldots)$
        \Else
            \State \textbf{restart}
        \EndIf
    \EndWhile
    \State \textit{Sort $Y$ to determine if its elements are strictly ordered}
\EndWhile
%\State $\sigma \gets \text{argsort}(Y)$
\State \Return the alternating sequence associated with $Y$
\end{algorithmic}
\end{algorithm}

This algorithm is essentially a rewritten version of Marchal's algorithm, modified to bypass inefficient trigonometric calculations. More precisely, by making the following change of variable $C_n = \sin^2\left(\frac{\pi}{2} X_n\right)$
where $X_n$ is the sequence described in \cite{Marchal2012}, we do not alter the properties of the sequence and we significantly improve the computation. We refer the reader to the initial note by Marchal \cite{Marchal2012} for the proof of his algorithm's validity. A key point is that the algorithm uses a rejection procedure but it is proved in \cite{Marchal2012} that the rejection probability is bounded above by $1 -\frac{2}{3\pi}$ (which is independent of $n$).

The remaining issue concerns the random-bit complexity. Naturally, for each uniform i.i.d. random variable \(U_i\), we aim to generate only enough digits to ensure sufficient precision, thus avoiding any order ambiguity in the final sequence.

As a preliminary, the number of strictly increasing binary trees of size \(n\) is asymptotically equivalent to \(2\left(\frac{2}{\pi}\right)^{n+1}n!\). Thus, the entropy is \(n \log_2(n)+\frac{2n}{\pi}+o(n)\). As we draw \(n\) uniform variables, we need at least \(\log_2(n)\) bits of precision to be coherent with the entropy.

Now, note that \(D[n]\) converges when $n$ goes to the infinity to the distribution
\[Q:=\sum_{n=0}^{\infty}(-1)^n\prod_{i=0}^{n}U_i^2\]

Indeed, we can derive from the algorithm the following induction: $Q=U^2(1-V^2Q)$
 where $U$ and $V$ are i.i.d. uniform distributions on $[0,1]$.
 Consequently, the density function \(d(z)\) of \(Q\) follows the functional equation:
\[d(z) = \int_{z}^{1} \frac{1}{2\sqrt{t}} \cdot \frac{d(1 - z/t)}{t} \, dt.\]
To our knowledge, solving this equation is not straightforward.
For that, let us start from the density of the random variable \(Q_0 = U^2(1-V^2)\), which is:
\[d(z) := \frac{\arctan\left(\frac{\sqrt{1 - z}}{\sqrt{z}}\right)}{2\sqrt{z}},\]
from which we can observe that the iteration \(Q_{n+1} = U^2(1-V^2Q_n)\) generates random variables whose density is alternatively a polynomial in \(A := \arctan\left(\frac{\sqrt{1 - z}}{\sqrt{z}}\right)\) and in \(\bar{A} := \arctan\left(\frac{\sqrt{z}}{\sqrt{1-z}}\right)\), divided by \(\sqrt{z}\). Moreover, the polynomials follow a quite simple recurrence which alternates
\[P_{n+1}(\bar{A}) = \int_{\bar{A}}^{\frac{\pi}{2}} P_n(A) \, dA\] and 
\[P_{n+1}(A) = \int_{0}^{A} P_n(\bar{A}) \, d\bar{A}.\]
So, the limiting polynomial follows the functional equation
\[\int_{0}^{x} \left( \int_{x}^{\frac{\pi}{2}} P(t) \, dt \right) dx = P(x),\]
from which we derive that the limiting distribution admits two solutions depending on the parity on $n$: \(\frac{2\sqrt{1 - z}}{\pi\sqrt{z}}\) and \(\frac{2\sqrt{z}}{\pi\sqrt{1-z}}\). Therefore, we can conclude that the density of the limiting random variable is
\[ds(z) = \frac{\left(\frac{\sqrt{1 - z}}{\sqrt{z}}\right) + \left(\frac{\sqrt{z}}{\sqrt{1 - z}}\right)}{\pi}.\]

In order to evaluate the required precision, let us assume that all the $D[i]$ are independent and identically distributed and follow the symmetrized distribution \(Q\) of density $ds(z)$
This statement is not strictly true; however, it is asymptotically acceptable. This acceptability arises because the dependency between \(D[i]\) and \(D[j]\) decreases exponentially fast as the distance \(|j-i|\) increases. This property follows directly from \[Q:=\sum_{n=0}^{\infty}(-1)^n\prod_{i=0}^{n}U_i^2\] and is really important because it implies that the errors do not accumulate.

Let \(\delta\) denote the number of digits of precision (in base 2). Then, the probability \(P_{n,\delta}\) that there is no ambiguity in a calculation of the alternating sequence of size \(n\) with \(\delta\) digits of precision is given by:
\[
P_{n,\delta} := n![z^n]\prod_{i=1}^{2^\delta-1}\left(1+z\frac{ds\left(\frac{i}{2^\delta}\right)}{\sum_{i=1}^{2^\delta-1} ds\left(\frac{i}{2^\delta}\right)}\right),
\]
Now, putting \(N:=2^\delta\) and \(p_{i,n}:=\frac{ds\left(\frac{i}{N}\right)}{\sum_{i=1}^{N-1}ds\left(\frac{i}{N}\right)}\), when \(n\) is large, \(\sum_{i=1}^{N-1} ds\left(\frac{i}{N}\right)/N\) tends to \(1\), and \(\sum_{i=1}^{N-1} p_{i,n}^2\) is asymptotically equivalent to \(\frac{1}{n} \int_{1/n}^{1-1/n}ds(x)dx\), which simplifies to \(\sum_{i=1}^{N-1} p_{i,n}^2 \sim \frac{2\ln(n)}{\pi^2n}\). Consequently, \(P_{n,\delta}\) is asymptotically equivalent to \(1-\frac{N(N-1)}{2}\frac{2\ln(n)}{\pi^2n}\). Indeed, the extraction $[z^n]$ is nothing more than the $PSet_n$ of the $B(\boldsymbol{p})=\sum p_{i,n}$ where the $p_{i,n}$ are seen as $n-1$ different atoms and it is well known that $PSet_n$ can be expressed as a multivariate polynomial in the $B(\boldsymbol{p}^i)$. Moreover the two dominant contributions are $\frac{B(\boldsymbol{p})^n}{n!}-\frac{B(\boldsymbol{p}^2)B(\boldsymbol{p})^{n-2}}{2(n-2)!}$. Finally, to achieve a probability of rejection \(\varepsilon\), we require \(\delta = 2\log_2(n) + \log_2(\ln(n))+\kappa+\log_2{\varepsilon}\).
Thus, the required precision is proportional (factor 2) to the entropy. It is worth noting that improving the precision by one unit halves the risk of ambiguity.

This algorithm is highly efficient, but it cannot be extended to other types of increasing trees. The following section introduces a more flexible approach.

\section{Toward an Efficient Recursive Methods Without Preliminary Calculations}

We aim at generating a random strictly increasing binary tree with an exact size of $n$. The generating function for our combinatorial family satisfies the non-linear differential equation $T'(z) = T^2(z) + 1$ and let us denote by $t_n$ the number of trees of size $n$. Utilizing the symbolic method for combinatorial structures, we derive a recursive method algorithm for our purpose. Our approach is now to avoid the exact and costly computation of $t_n$ by a well-controlled approximation process, which paves the way for a systematic and efficient algorithm to generate these structures, as shown in Algorithm \ref{algo:recursive}.

\begin{algorithm}
    \caption{Generating $T$ of Size $n$ with the recursive method}
    \label{algo:recursive}
    \begin{algorithmic}[1]

        \State  \textbf{If $n=1$ :} Return the tree with one leaf labeled $1$
        \State \emph{Intelligently} generate $M \in \{ 0, 1 , .. , n-1 \}$ the size of the left son of $T$. Note that $P(M = m)$ has to be proportional to $\binom{n-1}{m}t_mt_{n-1-m}$.
        \State Generate recursively $T'$ and $T''$ of size $m$ and $n-1-m$ and return $T$ with $T'$ as a left son and $T''$ as a right son with the indexes "shuffled" among the $n-1$ remaining atoms.
    \end{algorithmic}
\end{algorithm}

There are a few problems with this generation scheme: How do we compute the $t_n$'s? The classical way to do so is to pre-compute all of them but doing so requires at least $\Omega(n^2)$ in space and even more in time. Generating $M$ can also be an issue, the classical way to generate $M$ would require computing all the $t_n$'s and inverting the probability function which can be ineffective.

We describe a way to compute only the necessary bits of precision of $t_n$ and a way to generate $M$ without the exact knowledge of $t_n$, solving the two issues.

\section{Generating $M$, the size of the left son}

\subsection{Having a direct expression for $t_n$}

The differential equation $T'(z) =T^2(z) + 1$ can be integrated in $T(z) = \tan(z)$ because $T(0) = 0$ (there is no strictly increasing binary tree of size $0$).

We can hence use the tangent development: $\tan(z) = \sum_{k\geq 1} \frac{4^k(4^k-1)|B_{2k}|}{(2k)!}z^{2k-1}$ with $|B_{2k}| = \frac{2(2k)!\zeta(2k)}{(2\pi)^{2k}}$ being the absolute value of the Bernoulli numbers, and $\zeta$ being the Riemann's zeta function. So we can extract the number of strictly increasing binary trees of size $n$. If $n$ is even, $t_n = 0$ because $\tan$ is odd, and for $n = 2k -1$, $t_{2k-1} = (2k-1)!\frac{2(4^k-1)\zeta(2k)}{\pi^{2k}}$.

In the following, we assume that $n = 2l-1$ is odd since there is nothing to generate if $n$ is even.

Since $P(M = m)$ is proportional to $\binom{n-1}{m}t_mt_{n-m-1}$, so $P(M = m) = 0$ when $m$ is even. So we can write, for $m = 2k-1$, that $P(M = m)$ is proportional to $\binom{2l}{2k-1}\frac{(2k-1)! 2(4^k-1)4^k\zeta(2k)}{(2\pi)^{2k}}\frac{(2(l-k)-1)!(4^{l-k}-1)4^{l-k}\zeta(2(l-k))}{(2\pi)^{2(l-k)}}$. We can simplify this expression and get rid of the constant terms, we get that $P(M = 2k-1)$ is proportional to $f_k:= (4^k-1)\zeta(2k)(4^{l-k}-1)\zeta(2(l-k))$, with $k\in \{1,2,..,l-1\}$.

Since $\zeta(2k) = \sum_{i\geq 1} \frac{1}{i^{2k}}$, by truncating the sum, $\zeta(2k) \geq 1 + \frac{1}{4^k}$. This leads to the following bound on $f_k$: $f_k \geq (4^k-1)(1 + \frac{1}{4^k})(4^{n-k}-1)(1 + \frac{1}{4^{n-k}})$, i.e. $f_k \geq (4^k-\frac{1}{4^k})(4^{n-k}-\frac{1}{4^{n-k}})$.
With this lower bound and the fact that $(f_k)$ are symmetric decreasing from $1$ to $\lfloor \frac{l}{2} \rfloor$, we can bound $f_k$.

\begin{lemma}\label{bound f}
    For $k\in \{1,2,..,l -1 \}$, $f_1 \geq f_k \geq f_{\lfloor \frac{l}{2} \rfloor}$ and $\frac{f_1}{f_{\lfloor \frac{l}{2} \rfloor}} \leq \frac{4\pi^4}{45} \leq 9$
\end{lemma}

This lemma implies a Monte-Carlo method for sampling $M$.

\begin{algorithm}
\caption{Generate $M$ such that $P(M = 2k-1)$ is proportional to $f_k$}
\label{algo:MonteCarlo}
\begin{algorithmic}[1]
    \State Generate $X \in \{1,2,..,l - 1\}$.
    \State Test $U f_1 \leq f_{X}$ where $U$ is a uniform random number in $[0,1]$, if true return $M = X$, else generate another $X$ and start again.

\end{algorithmic}
\end{algorithm}

Lemma \ref{bound f} implies that Algorithm \ref{algo:MonteCarlo} has a constant reject.

\begin{lemma}
   Algorithm \ref{algo:MonteCarlo} has a running time of $O(\log(n))$
\end{lemma}

This lemma is not only due to the fact that the reject is constant but also to the fact that it is possible to compute only the first bits of $f_1$, $f_k$, and $U$ and check whether we can conclude that $U f_1 \leq f_{X}$ or $U f_1 > f_{X}$. If we cannot conclude, we can add the next bits and recheck until we can accept or reject $X$ as $M$.

It is possible to compute only the first bits of $f_k = (4^k-1)\zeta(2k)(4^{l-k}-1)\zeta(2(l-k))$ because we can truncate $\zeta(2k) = \sum_{i\geq 1} \frac{1}{i^{2k}}$ to a finite order.

With this method of generating $M$, we can state the complexity of Algorithm \ref{algo:recursive}.

\begin{theorem}
   Algorithm \ref{algo:recursive} has a running time of $O(n\log(n))$, which is optimal.
\end{theorem}

The algorithm is optimal because of the Shannon entropic principle: It is not possible to generate a strictly increasing binary tree of size $n$ ($n$ odd) with less than $\log_2(t_n)$ random bits, so the best sampling algorithm will run in at least $\log_2(t_n) = \Omega(n\log(n))$

\section{Adapting the Algorithm for Different Families of Binary-Increasing Trees}

\subsection{Binary Trees}

Consider various families of increasing binary trees, each characterized by distinct generating functions; for instance, non-plane binary trees have $T'(z) =\frac{T^2(z)}{2} + 1$, leading to $T(z) = \frac{\tan(z/\sqrt{2})}{\sqrt{2}}$. For all families of binary-increasing trees with generating functions of the form $T' = aT^2 + c$ as described in Algorithm \ref{algo:recursive}, the algorithm remains applicable. This is due to the consistent structure shared by these trees, characterized by having one more leaf than internal nodes.

\subsection{Binary-Unary Trees}

For a family with generating function $T'(z) = a(T(z)+ \alpha)^2 + a\gamma^2$, where $a,c>0$ and $b\geq 0$, we focus on those written as $T'(z) = a(T(z)+ \alpha)^2 + a\gamma^2$, the case $T'(z) = a(T(z)+ \alpha)^2 - a\gamma^2$ being similar. The algorithm generates trees of size $n$ with $t_n \sim \frac{n!a^n\gamma^{n+1}}{(\frac{\pi}{2} - \tan^{-1}(\frac{\alpha}{\gamma}))^{n+1}}$, allowing effective recursive generation. And $$t_n = n!a^n\gamma^{n+1}\left( \sum_{k\geq 0} \frac{(-1)^{n+1}}{(\tan^{-1}(\frac{\alpha}{\gamma}) - \pi(k+\frac{1}{2}))^{n+1}} + \frac{(-1)^{n+1}}{(\tan^{-1}(\frac{\alpha}{\gamma}) + \pi(k+\frac{1}{2}))^{n+1}} \right)$$ is still computable by truncating its expression.

\begin{algorithm}
\caption{Generating $T$ of Size $n$ for Binary-Unary Trees}
\label{algo:recursivev2}
\begin{algorithmic}[1]

    \State  \textbf{If $n=1$ :} Return the tree with one leaf colored in one of the  $c$ colors.
    \State \textbf{For $n>1$ :} The tree can have a unary root with probability $\frac{bt_{n-1}}{t_n}$ or a binary root.
    \State Let $X$ be a Bernoulli variable with parameter $\frac{bt_{n-1}}{t_n}$; if $X = 1$, call the algorithm recursively with parameter $n-1$ and return the tree consisting of a unary root and labeled $1$ with $T'$ as its only son.
    \State Generate $M \in \{ 0, 1 , .. , n-1 \}$ with $P(M = m)$ proportional to $\binom{n-1}{m}t_mt_{n-1-m}$.
    \State Generate recursively $T'$ and $T''$ of size $m$ and $n-1-m$ and return $T$ with $T'$ as a left son and $T''$ as a right son with the indexes "shuffled" among the $n-1$ remaining axioms.
\end{algorithmic}

\end{algorithm}

\bibliographystyle{eptcs}
\bibliography{generic}

\begin{thebibliography}{1}
\providecommand{\bibitemdeclare}[2]{}
\providecommand{\surnamestart}{}
\providecommand{\surnameend}{}
\providecommand{\urlprefix}{Available at }
\providecommand{\url}[1]{\texttt{#1}}
\providecommand{\href}[2]{\texttt{#2}}
\providecommand{\urlalt}[2]{\href{#1}{#2}}
\providecommand{\doi}[1]{doi:\urlalt{https://doi.org/#1}{#1}}
\providecommand{\eprint}[1]{arXiv:\urlalt{https://arxiv.org/abs/#1}{#1}}
\providecommand{\bibinfo}[2]{#2}

\bibitemdeclare{article}{BBJ17}
\bibitem{BBJ17}
\bibinfo{author}{Axel \surnamestart Bacher\surnameend},
  \bibinfo{author}{Olivier \surnamestart Bodini\surnameend} \&
  \bibinfo{author}{Alice \surnamestart Jacquot\surnameend}
  (\bibinfo{year}{2017}): \emph{\bibinfo{title}{Efficient random sampling of
  binary and unary-binary trees via holonomic equations}}.
\newblock {\slshape \bibinfo{journal}{Theor. Comput. Sci.}}
  \bibinfo{volume}{695}, pp. \bibinfo{pages}{42--53},
  \doi{10.1016/J.TCS.2017.07.009}.

\bibitemdeclare{article}{DFLS004}
\bibitem{DFLS004}
\bibinfo{author}{Philippe \surnamestart Duchon\surnameend},
  \bibinfo{author}{Philippe \surnamestart Flajolet\surnameend},
  \bibinfo{author}{Guy \surnamestart Louchard\surnameend} \&
  \bibinfo{author}{Gilles \surnamestart Schaeffer\surnameend}
  (\bibinfo{year}{2004}): \emph{\bibinfo{title}{Boltzmann Samplers for the
  Random Generation of Combinatorial Structures}}.
\newblock {\slshape \bibinfo{journal}{Combinatorics, Probability and
  Computing}} \bibinfo{volume}{13}(\bibinfo{number}{4–5}), p.
  \bibinfo{pages}{577–625}, \doi{10.1017/S0963548304006315}.

\bibitemdeclare{unpublished}{Marchal2012}
\bibitem{Marchal2012}
\bibinfo{author}{Ph. \surnamestart Marchal\surnameend} (\bibinfo{year}{2012}):
  \emph{\bibinfo{title}{Generating random alternating permutations in time $n
  \log n$}}.
\newblock
  \urlprefix\url{https://www.math.univ-paris13.fr/~marchal/altperm1.pdf}.
\newblock \bibinfo{note}{Unpublished note}.

\end{thebibliography}
\end{document}